\documentclass[12pt]{iopart}
\usepackage{iopams}  
\usepackage{graphicx}  
\usepackage{color}

\newcommand{\pbar}         {$\rm\overline{p}$}
\newcommand{\kpim}          {${\rm K}^{-}/\pi^{-}$}
\newcommand{\ppim}          {$\bar{{\rm p}}/\pi^{-}$}

\newcommand{\dndy}         {d$N$/d$y$}

\newcommand{\dNdeta}       {\ensuremath{\mathrm{d}N_\mathrm{ch}/\mathrm{d}\eta}}

\newcommand{\s}            {\ensuremath{\sqrt{s}}}
\renewcommand{\pt}           {\ensuremath{p_{\rm t}}}
\newcommand{\meanpt}           {\ensuremath{\langle p_{\rm t} \rangle}}

\newcommand{\snn}          {\ensuremath{\sqrt{s_{\rm NN}}}}

\newcommand{\kzero}          {\ensuremath{{\rm K}^{0}_{S}}}

\newcommand{\betaT}        {\ensuremath{\langle \beta_{\rm t} \rangle}}

\begin{document}

\title[Identified particles in pp and Pb-Pb collisions with the ALICE detector]{Identified particles in pp and Pb-Pb collisions at LHC energies with the ALICE detector}

\author{M Floris for the ALICE Collaboration\footnote{The full list of
    authors can be found at the end of this volume.}}

\address{CERN, Geneva, Switzerland}
\ead{michele.floris@cern.ch}
\begin{abstract}
The ALICE experiment has been taking  data since 2009, with proton and lead beams. In this paper, the different particle identification techniques used by the experiment are briefly reviewed. The current results on identified particle spectra in pp collisions at $\sqrt{s} = 900~\textrm{GeV}$ and 7~TeV, and in Pb-Pb collisions at  $\snn = 2.76~\mathrm{TeV}$ are summarized. In particular, the energy dependence of the spectral shapes and particle ratios in pp collisions is discussed and the results are compared to previous experiments and commonly used Monte Carlo models. The baryon/meson ratio $\Lambda/K^0_S$ is studied in Pb-Pb collisions as a function of transverse momentum and centrality, and it is compared to previous results. The evolution of the particle spectra in Pb-Pb with collision centrality is compared to measurements at lower energies and discussed in the context of thermal and hydrodynamical models. 
\end{abstract}


\section{Introduction}
\label{sec:introduction}

The ALICE experiment has unique particle identification (PID) capabilities among the LHC experiments. Nearly all known techniques are employed, allowing the experiment to identify a large variety of particles over an extended range in transverse momentum. 
The measurement of identified particle spectra is a crucial ingredient in the understanding of heavy ion collisions, as it allows to access the thermal parameters of the system at freeze-out and it poses strict constraints on the (hydrodynamic) models aiming to describe the data. The study of identified particles in pp collisions provides valuable reference spectra for the understanding of heavy ion data. Moreover, pp results have their own genuine interest in the context of minimum bias physics, underlying event and the tuning of Monte Carlo generators.
In this paper,  we first review the PID detectors relevant to the present discussion (sec.~\ref{sec:part-ident-alice}). New results obtained in pp collisions at \s~=~7~TeV (sec.~\ref{sec:results-pp-collsions}) and Pb-Pb
collisions at \snn~=~2.76~TeV (sec.~\ref{sec:results-pb-pb}) are then presented. The results are finally summarized in sec.~\ref{sec:conclusions}. In this work, we present particle ratios and spectra for primary particles, defined as prompt particles produced in the collision and all decay products, except products from weak decays of strange particles.

\section{Particle Identification in ALICE}
\label{sec:part-ident-alice}

\begin{figure}[tbp]
  \centering
  \includegraphics[width=0.49\textwidth]{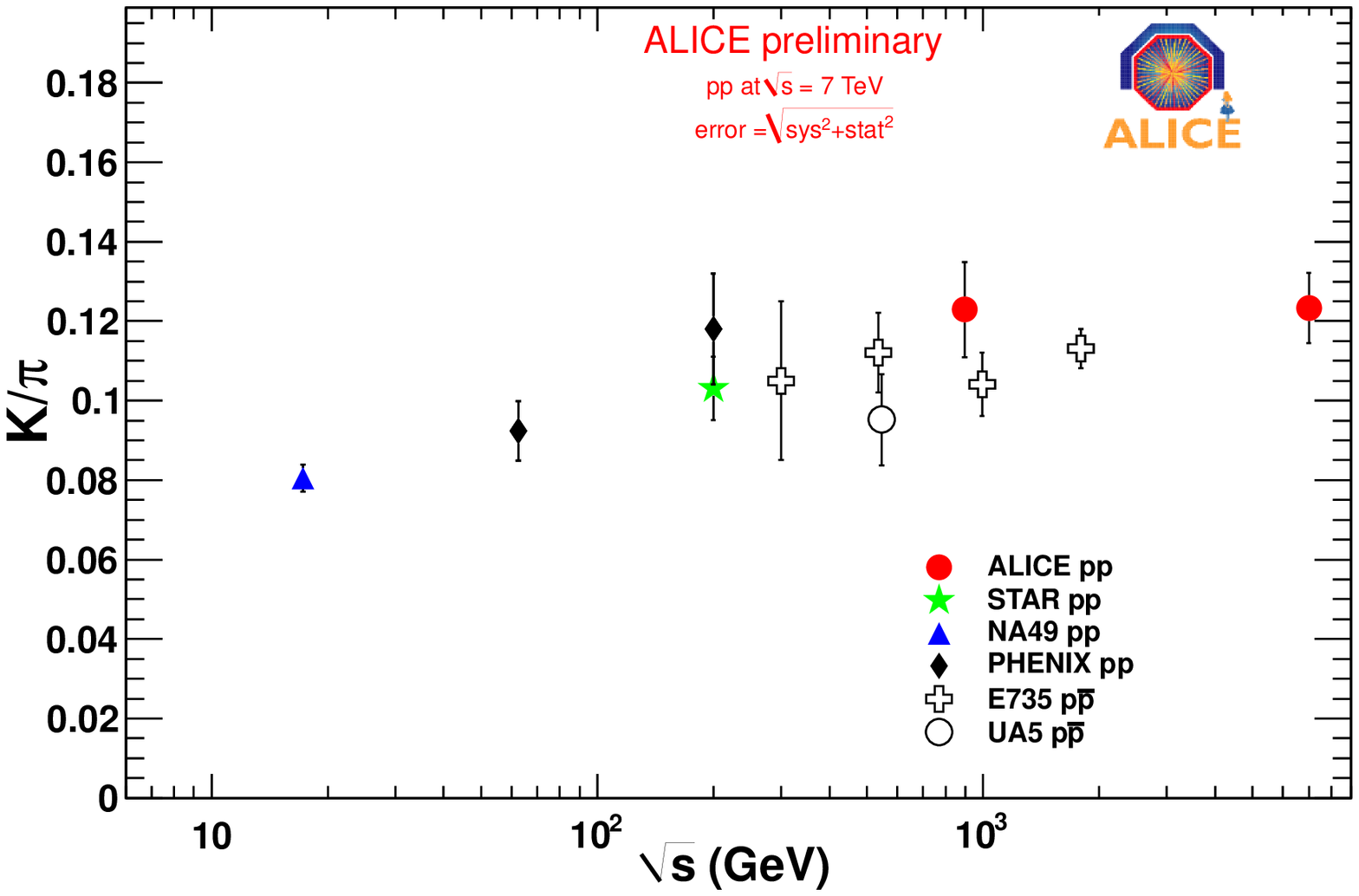}
  \hfill
  \includegraphics[width=0.49\textwidth]{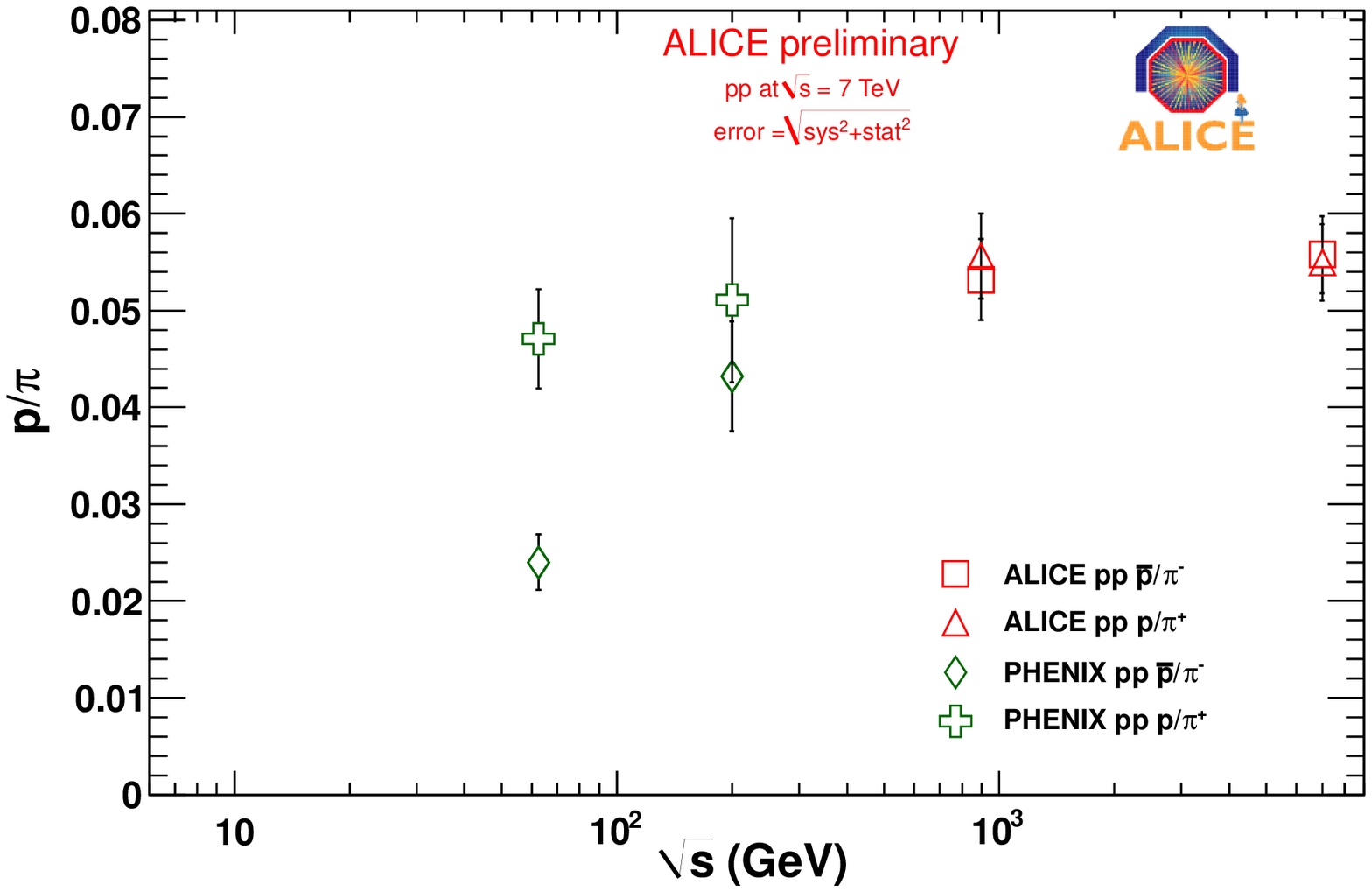}
 \caption{($\mathrm{K}^++\mathrm{K}^-$)/($\pi^{+}+\pi^{-}$) (left panel) and p/$\pi^{+} $, $\bar{\mathrm{p}}/\pi^{-}$ (right panel) as a function of \s.}
  \label{fig:kpi-and-ppi-ratios}
\end{figure}

The PID capabilities of the ALICE experiment are discussed in detail in~\cite{Alexander-QM,Alessandro:2006yt,Aamodt:2008zz}. In this section, we briefly review the detectors relevant for the present analyses. 

Particles are reconstructed  close to the interaction point with the Inner Tracking System (ITS), which is composed of 6 layers of silicon detectors, utilizing different technologies. This detector can provide identification via the specific energy loss in 4 of the 6 layers. The ITS can also work as a standalone tracker, allowing the reconstruction of low \pt\ tracks, which do not have enough momentum to reach the outer trackers. This option makes the identification of $\pi$, K, p (``stable hadrons'') possible down to 0.1, 0.2, 0.3~GeV/$c$ respectively.
The ITS is followed by a large volume Time Projection Chamber (TPC), the main tracking device in the experiment. The tracks reconstructed in the TPC can be combined with the information from the ITS to form ``global tracks'', which provide better resolution in the distance of closest approach to the vertex, and hence better separation of primary and secondary particles. This is the case of most analyses presented in this paper. The TPC can identify particles  via the specific energy loss in the fill gas: up to 159 samples can be measured. 
Further outwards, the Time of Flight (TOF) detector measures the arrival time of the particles, allowing identification at higher \pt\ than accessible with the TPC. The total time resolution is determined by the intrinsic time resolution of the TOF detector and by the start time resolution. The latter depends on the multiplicity of the event, and hence on the colliding system. The total resolution is about 85~ps in Pb-Pb and 120~ps in pp collisions.

Particles can also be identified in ALICE with topological identification or invariant mass fits. This is the case for weak decays, resonances and decaying kaons (``kinks''). In those studies, the PID detectors can be used to improve the signal over background ratio, without any loss of the actual signal, by means of ``compatibility cuts'' with the PID signal (e.g. requiring that the dE/dx signals of the kaons in a $\phi \to \mathrm{KK}$ candidate are within 3 standard deviations from the expected average value).

\section{Results in pp collisions at \s~=~900~GeV and 7~TeV}

\label{sec:results-pp-collsions}

In this section we present new results for $\pi$, K, p, resonances and cascades measured in pp collisions at \s~=~7~TeV. The data are compared to existing measurements at lower energy, in particular published ALICE results at \s~=~900~GeV~\cite{Aamodt:2011zz,Aamodt:2011zj}.

The stable hadrons were measured combining the techniques and detectors described in sec.~\ref{sec:part-ident-alice}. More details can be found in~\cite{Marek-QM}. In Fig.~\ref{fig:kpi-and-ppi-ratios} we show the ratios K/$\pi$ and p/$\pi$ as a function of energy. As it can be seen, the K/$\pi$ ratio is rather independent of energy, at least starting from RHIC energies (\s~=~200 GeV), despite the large increase in the center of mass energy. The p/$\pi$ ratio is shown for separate charges in the right panel. The difference between the two charges at lower energies reflects the baryon/antibaryon asymmetry, which essentially vanishes at LHC energies as already reported in~\cite{Aamodt:2010dx},  leading to a constant value of about 0.05 for the two energies measured by ALICE.

Results for $\mathrm{K}^*(892)$, $\phi(1020)$ and $\Xi^*(1530)$ production in pp collisions at \s~=~7~TeV were reported at this conference~\cite{Alberto-QM,Alessandro-QM}. We would like to stress here that most of the ratios are seen to be constant over a wide energy range, in line with the above observation about the K/$\pi$ ratio.

\begin{figure}[tbp]
  \centering
  \includegraphics[height=0.39\textwidth]{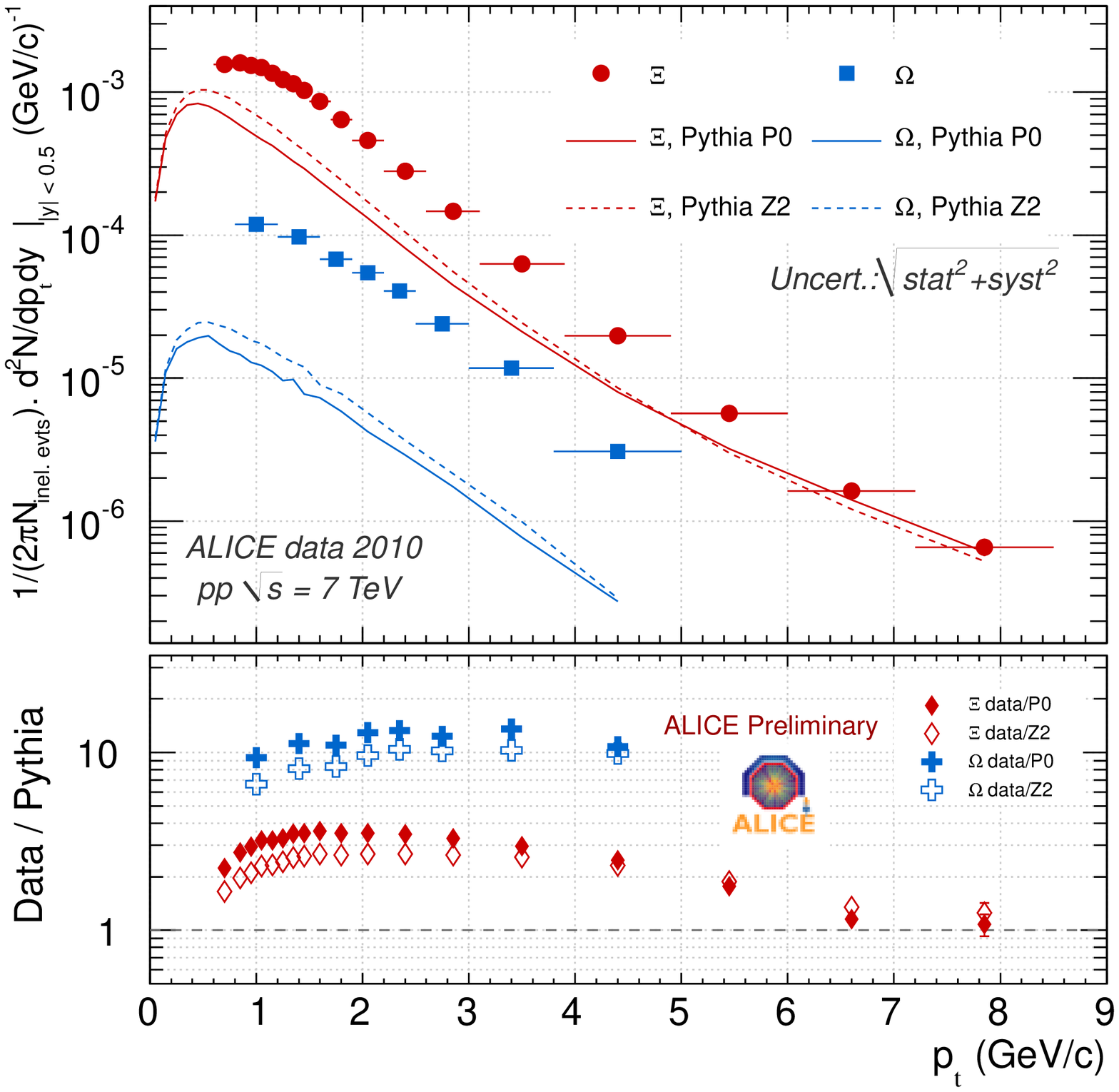}
  \hfill
  \includegraphics[height=0.39\textwidth]{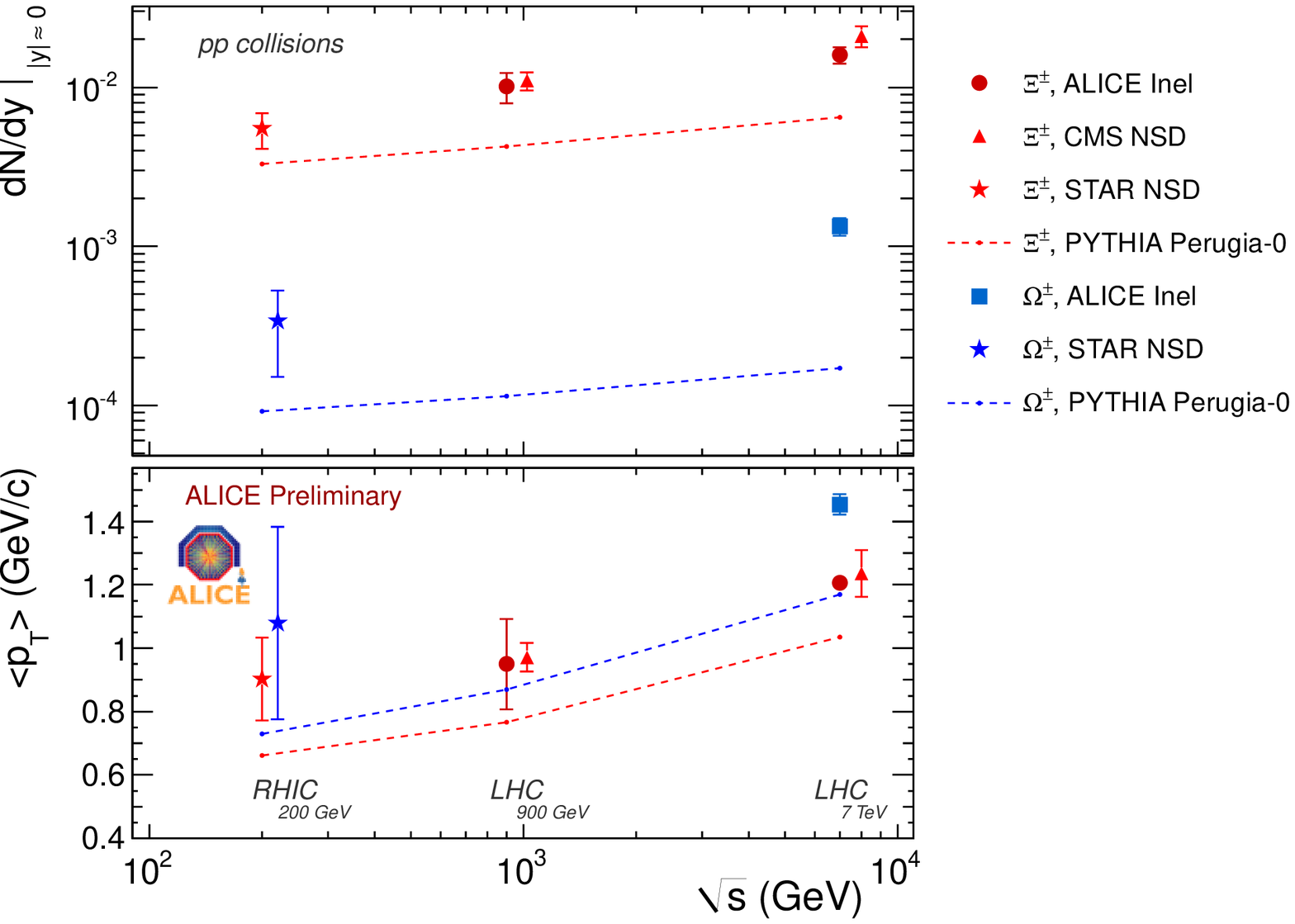}
 \caption{Left: $\Xi$ and $\Omega$ spectra compared to 2 tunes of the Pythia event generator. Right: \s-dependence of yields and \meanpt.}
  \label{fig:cascades}
\end{figure}

The multi-strange baryons $\Xi$ and $\Omega$  were measured via topological identification of the decays channels $\Xi\to\Lambda+\pi\to \mathrm{p} +\pi+\pi$ and $\Omega\to \Lambda + \mathrm{K} \to \mathrm{p} + \pi + \mathrm{K}$. In particular, this is the first measurement of the $\Omega$ at the LHC~\cite{David-QM}. The spectra are compared to recent tunes of the Pythia event generator~\cite{Sjostrand:2006za,Skands:2010ak,Field:2010bc} in Fig.~\ref{fig:cascades}. They significantly under-predict the data, by a factor of about 10 in the measured range for the $\Omega$. The shape of the \pt\ spectra is also not reproduced. This is a general feature of all strange particles, and can provide valuable input for the tuning of strange particle production in the Monte Carlo  generators. In the right panel of Fig.~\ref{fig:cascades}, we show the \s\ dependence of the mean \pt\ and of the \dndy, compared to previous measurements~\cite{Abelev:2006cs,Khachatryan:2011tm} and to the Monte Carlo models. The spectra become harder with increasing \s\ and the \meanpt\ of the $\Omega$ is significantly larger than that of the $\Xi$. 
The \meanpt\ and yield of the $\Xi$ measured at 7~TeV by ALICE and CMS are seen to be in agreement. For the yield, one has to bear in mind that the CMS result is normalized to non-single-diffractive events (NSD), while the ALICE results are normalized to inelastic collisions, and there is an overall $\sim$ 20\% difference between the two classes~\cite{Martin-QM}.

\begin{figure}[tbp]
  \centering
  \includegraphics[height=0.45\textwidth]{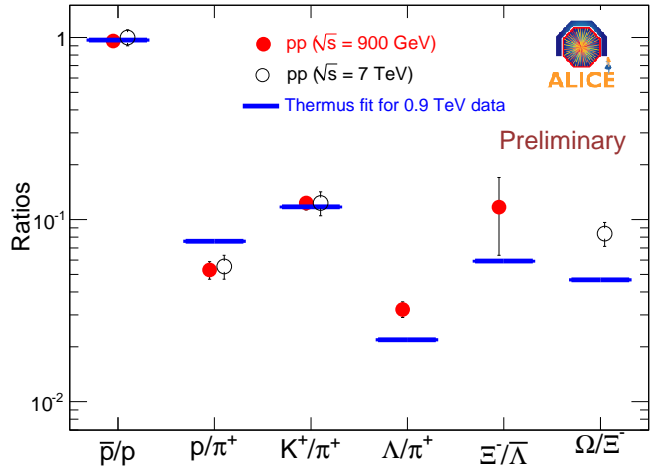}
  \caption{Thermal fit of the \s~=~900~GeV pp data, compared to new results at \s~=~7~TeV.}
  \label{fig:thermal-fit-pp}
\end{figure}

The statistical hadronization model was successfully used in the past to describe pp collisions at lower energies~\cite{Becattini:2010sk,Kraus:2008fh}.
The same kind of analysis was performed on the pp results at 900 GeV~\cite{Oeschler:2011ny}, as shown in Fig.~\ref{fig:thermal-fit-pp}, where the ratios are fitted with the THERMUS model~\cite{Wheaton:2004qb}. The thermal model yields a poor description of the data. The new 7~TeV results are also shown: the ratios are not changing between the two energies, leading to an equally poor description.  On the other hand, comparing the data with predictions for LHC energies~\cite{Kraus:2008fh,Becattini:2009ee}, it can be seen that most of the ratios (except those involving protons) agree with the predictions within $\sim 20\%$. In particular, the yields for multi-strange baryons is much closer to the predictions of the thermal model than to the Pythia tunes.

\section{Results in Pb-Pb collisions at \snn~=~2.76~TeV}
\label{sec:results-pb-pb}

In this section we address a few aspects of identified particle production in Pb-Pb collisions: the baryon meson anomaly through the $\Lambda/\kzero$ ratio, and the kinetic and chemical properties of the system through identified spectra of $\pi$/K/p.

The baryon/meson ratio in heavy ion collisions was seen at RHIC to be enhanced with respect to pp collisions~\cite{Lamont:2007ce}. The enhancement becomes stronger with increasing centrality. It features a maximum at intermediate $\pt \simeq 2.5~\mathrm{GeV}/c$, which is pushed towards larger \pt\ at higher centralities. These observations were understood in some models as a consequence of quark recombination (or ``coalescence'')~\cite{Greco:2003xt,Fries:2003vb}: in heavy ion collisions a quark can hadronize by picking up another quark from the medium. This increases the probability of forming a baryon at intermediate \pt, relative to mesons. The shift of the maximum towards higher \pt\ is interpreted in this language as a consequence of the larger radial flow for more central collisions.

\begin{figure}[tbp]
  \includegraphics[width=0.49\textwidth]{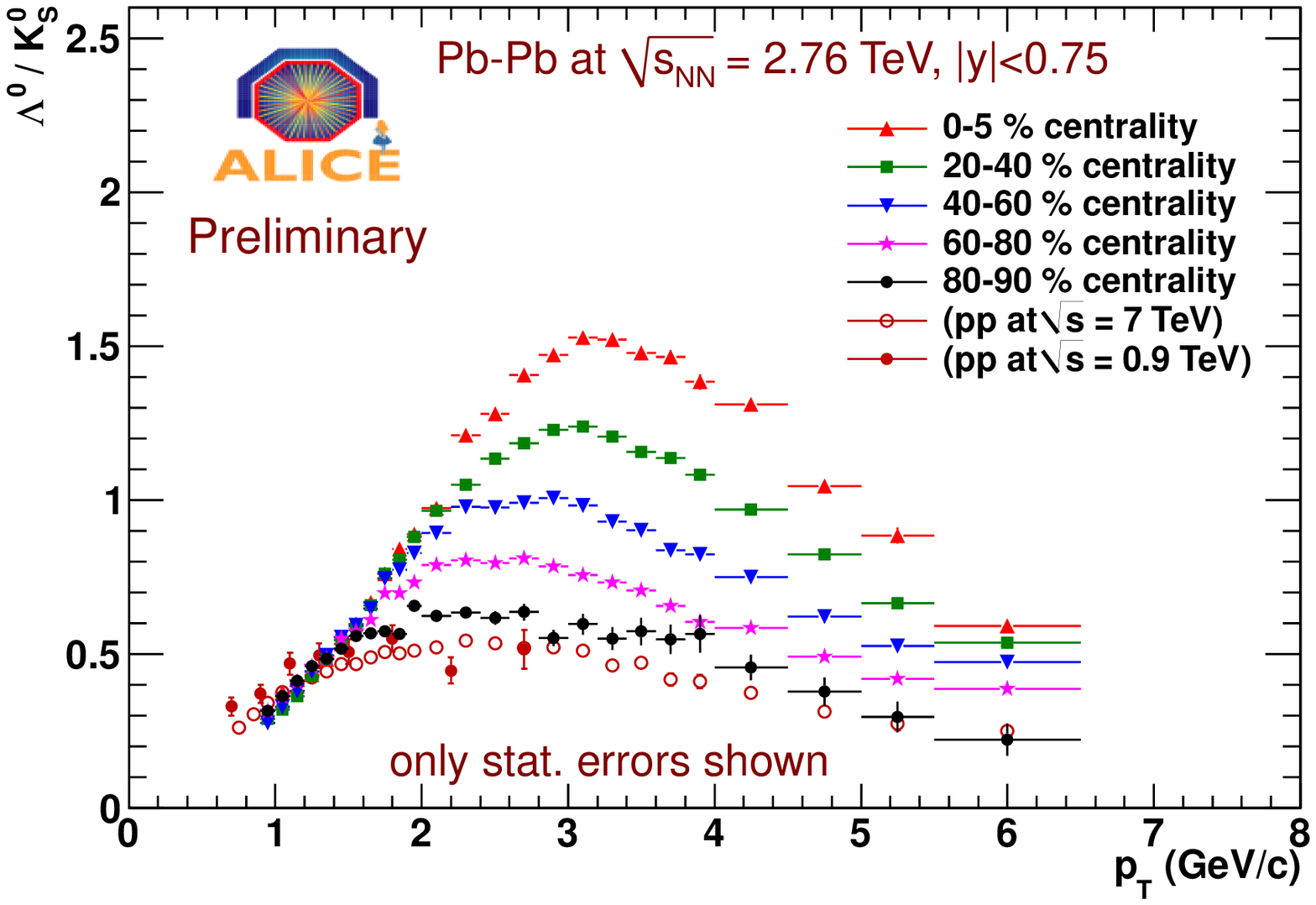}
  \centering
  \hfill
  \includegraphics[width=0.49\textwidth]{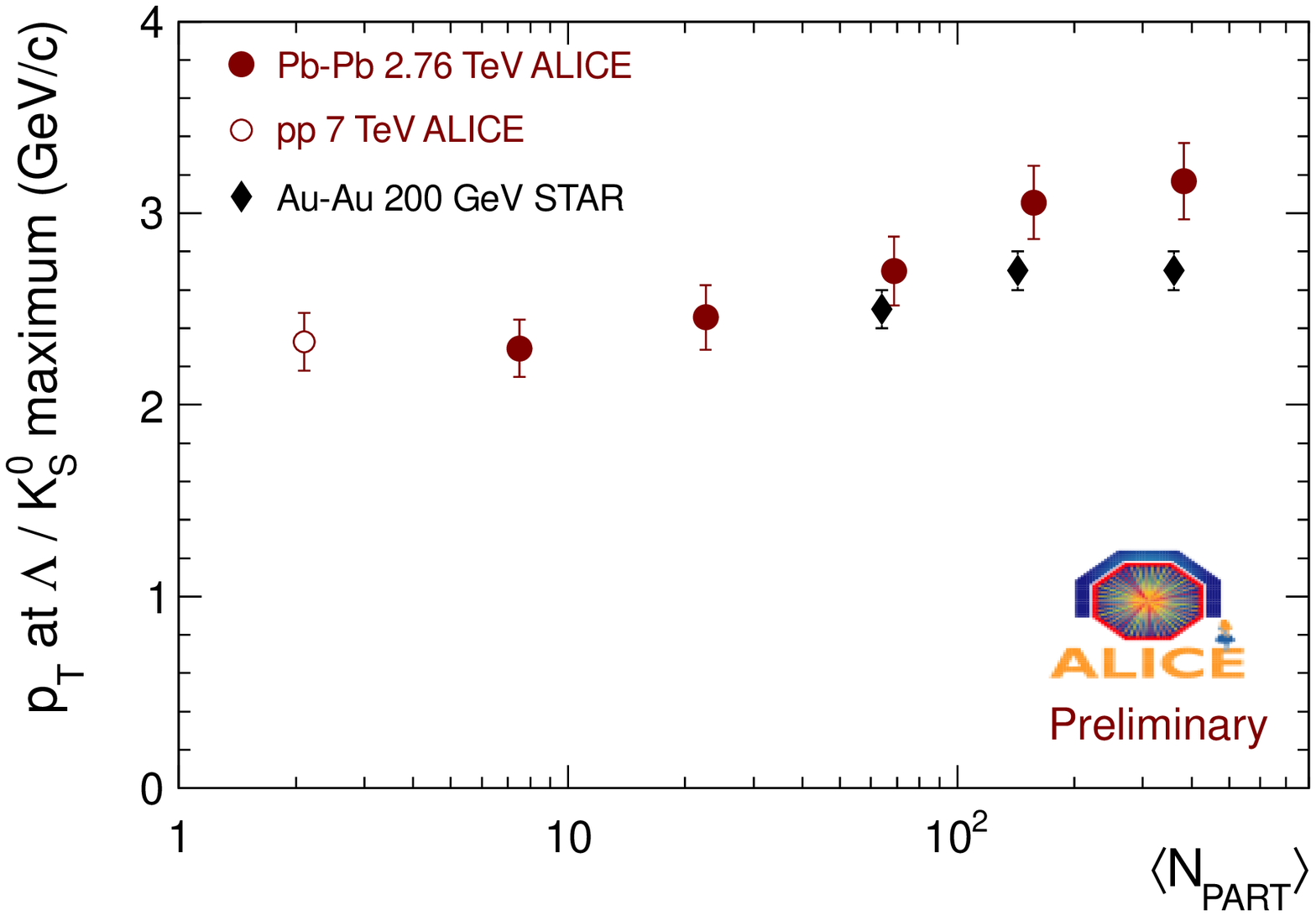}
 \caption{Left: $\Lambda/\kzero$ ratio in pp and Pb-Pb data as a function of \pt. Right: \pt\ at maximum of the ratio.}
  \label{fig:lambdak0}
\end{figure}
 
The ALICE measurement of the $\Lambda/\kzero$ ratio is shown in the left panel of Fig.~\ref{fig:lambdak0} for different centrality bins~\cite{Jouri-QM}. The ratio is enhanced with respect to pp collisions also at the LHC. When compared to RHIC results, the enhancement is observed to be slightly larger, with a maximum shifted to larger \pt\, as depicted in the right panel of Fig.~\ref{fig:lambdak0}. The enhancement is also observed to decrease less rapidly with \pt, being still a factor $\sim 2$ higher than at RHIC for $\pt \simeq 6~\mathrm{GeV}/c$~\cite{Jouri-QM}. The dramatic shift of baryon enhancement, of order
1-2 GeV, predicted by some models~\cite{Fries:2003fr} for LHC energies is not seen.

All the particle/antiparticle ratios are consistent with a value of one at the LHC, so we focus on negative spectra below. The spectra of $\pi$, K and p were measured using a combined ITS, TPC and TOF analysis, similar to what was done in pp (see also sec.~\ref{sec:results-pp-collsions}). 

\begin{figure}[tbp]
  \centering
  \includegraphics[width=0.49\textwidth]{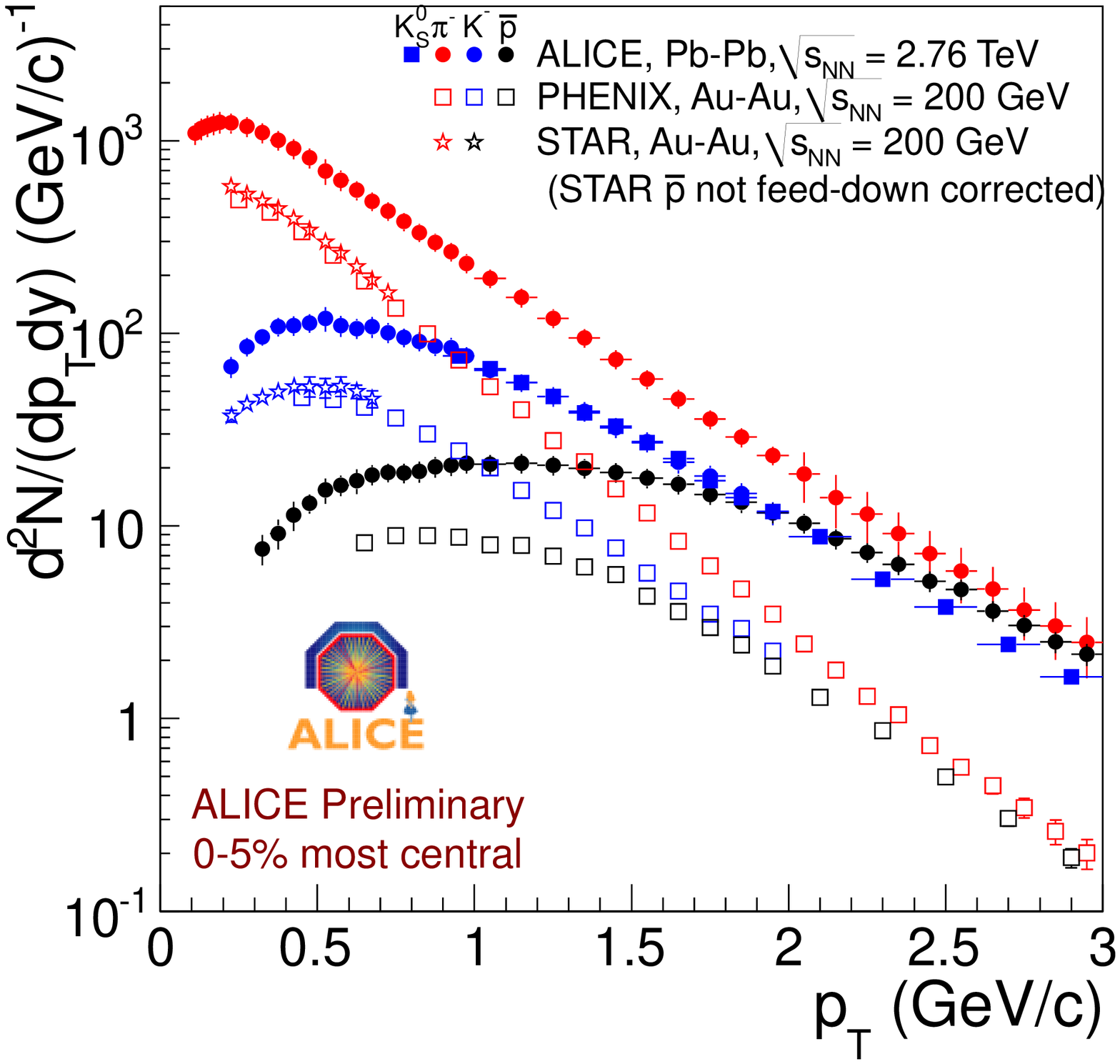}
  \hfill
  \includegraphics[width=0.49\textwidth]{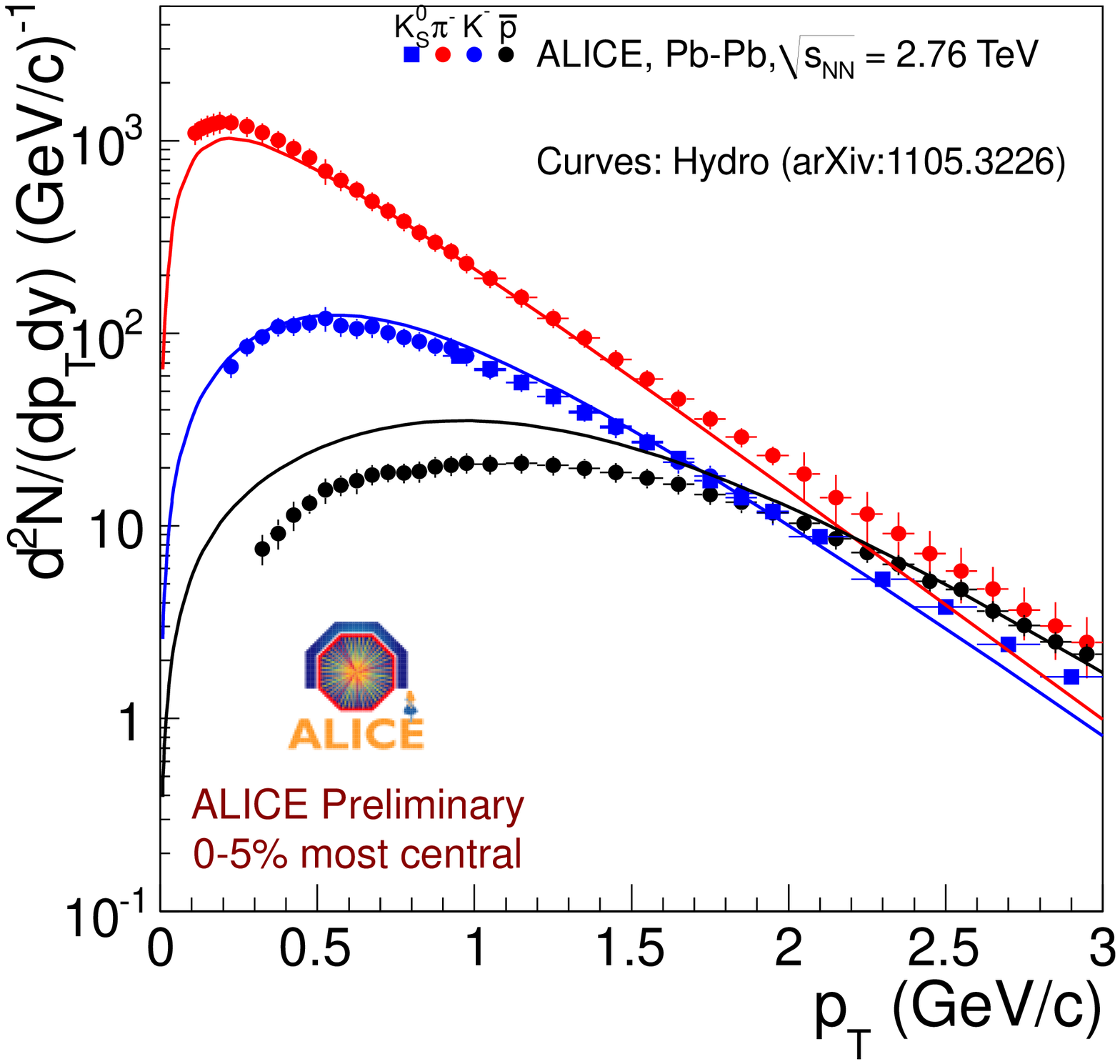}
 \caption{Spectra in the 0-5\% most central bin, compared to previous results at lower energy (left panel) and to the prediction from a hydrodynamical model (right panel).}
  \label{fig:centralPbPb}
\end{figure}

The spectra in the 0-5\% most central collisions are compared to previous results at \snn~=~200~GeV~\cite{:2008ez,Adler:2003cb} in the left panel of Fig.~\ref{fig:centralPbPb}. In order to extend the \pt\ range for kaons, the \kzero\ spectra are also shown on the same plot. The $\pi$ and p curves cross at $\pt \simeq 3~\mathrm{GeV/c}$, similar to what was observed in the $\Lambda/\kzero$ ratio.
The ALICE results refer to primary particles as defined in the introduction, i.e.~feed-down from weak decays are subtracted. At RHIC the situation is not always homogeneous: the (anti)protons from PHENIX are corrected for feed-down, while the ones from STAR are usually not. For this reason, only \pbar\ from PHENIX are shown in Fig.~\ref{fig:centralPbPb}. On the other hand, the $\pi$ from PHENIX are not corrected for feed-down, but since this is a much smaller correction than in the case of (anti)protons, this data are also presented in the figure. A dramatic change in shape is observed, with the spectra at the LHC being much flatter at low \pt\ and harder, indicating a stronger radial flow. In the right panel, the data are also compared to a hydrodynamical prediction~\cite{Shen:2011eg}. While the model catches the gross features of the $\pi^-$ and K$^-$ distribution, it strongly disagrees with the measured \pbar\ spectrum, both in shape and yield. The difference in shape indicates a stronger radial flow in the data than expected from the model, which could be partially due to extra flow built up in the hadronic phase~\cite{Song:2010aq}. A similar disagreement was seen when comparing the $\mathrm{v}_{2}$ of protons to the same model~\cite{Raymond-QM}. The difference in yield can be ascribed to the fact that the model uses yields from a thermal model with T~=~165~MeV, and this is found not to agree with our data (see below).

\begin{figure}[tbp]
  \centering
 \includegraphics[height=0.4\textwidth]{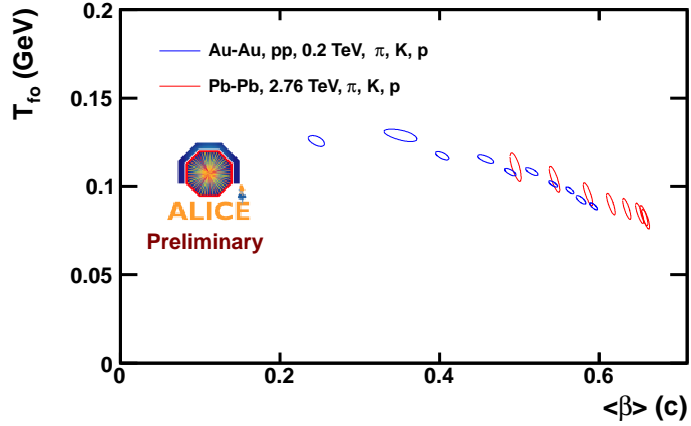}
 \caption{$T_{fo}$ and \betaT\ extracted from a blast wave fit, compared to previous results at \s~=~200~GeV.}
  \label{fig:blast-wave}
\end{figure}

In order to quantify the freeze-out parameters at \snn~=~7~TeV, we performed a  combined fit of our spectra with the blast wave function~\cite{Schnedermann:1993ws}. $\pi$, K, p spectra were fitted in the ranges 0.3 - 1 GeV/$c$, 0.2 - 1.5 GeV/$c$, 0.3 - 3 GeV/$c$ respectively, as the $\pi$ at low \pt\ are known to have a large contribution from resonance decays, while at high \pt\ a hard contribution (not expected to be described by the blast wave) may set in. It should be noticed that the value of the $T_{fo}$ parameter extracted from the fit is sensitive to the fit range used for the pions, because of the feed-down from resonances. This effect  will be studied in detail in the future. The results for the different centrality bins are compared in Fig.~\ref{fig:blast-wave} with similar fits made by the STAR collaboration~\cite{Adams:2005dq}. We observe a $\sim$10\% higher radial flow for most central collisions at the LHC. 

\begin{figure}[tbp]
  \centering
  \includegraphics[width=0.49\textwidth]{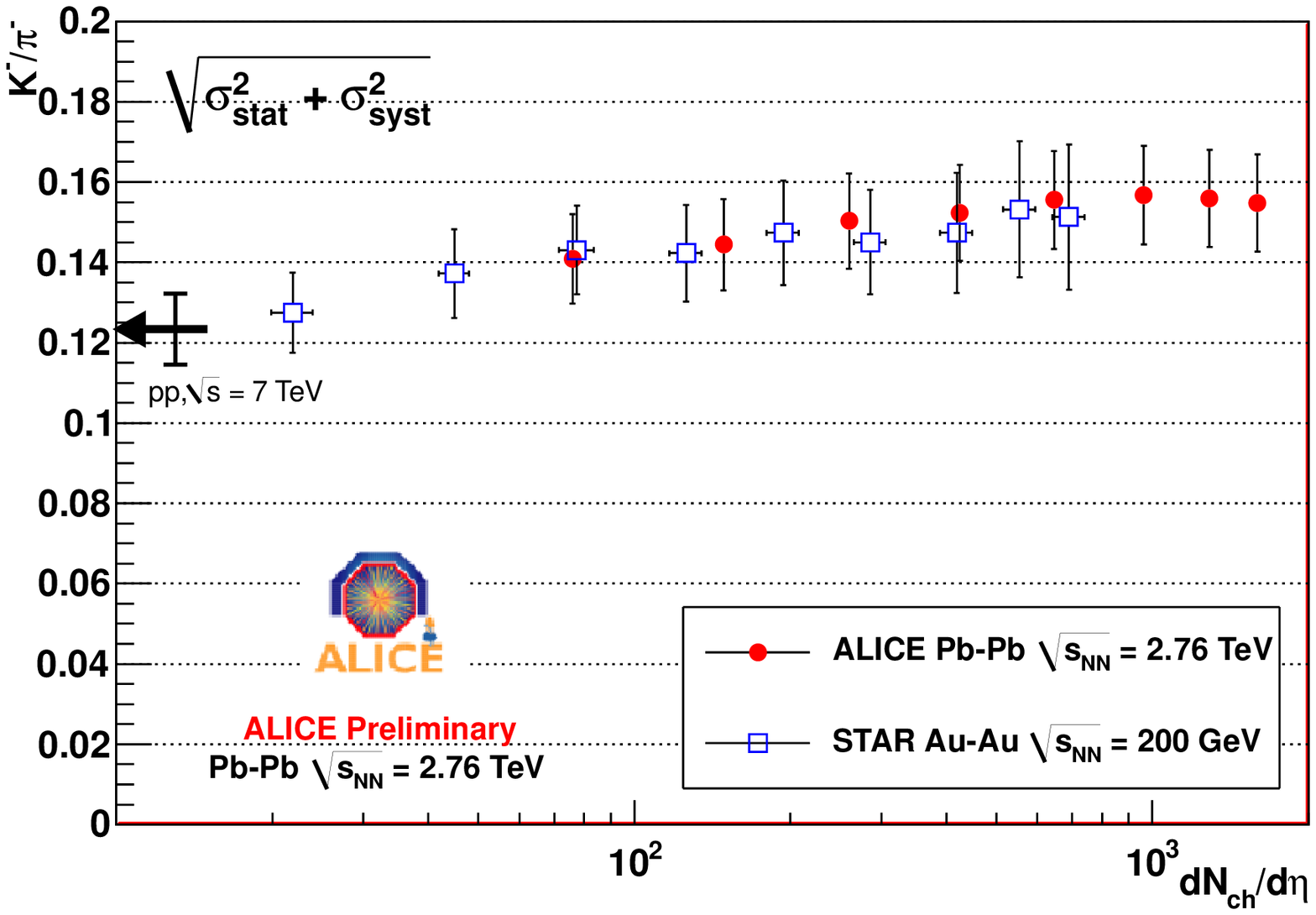}
  \hfill
  \includegraphics[width=0.49\textwidth]{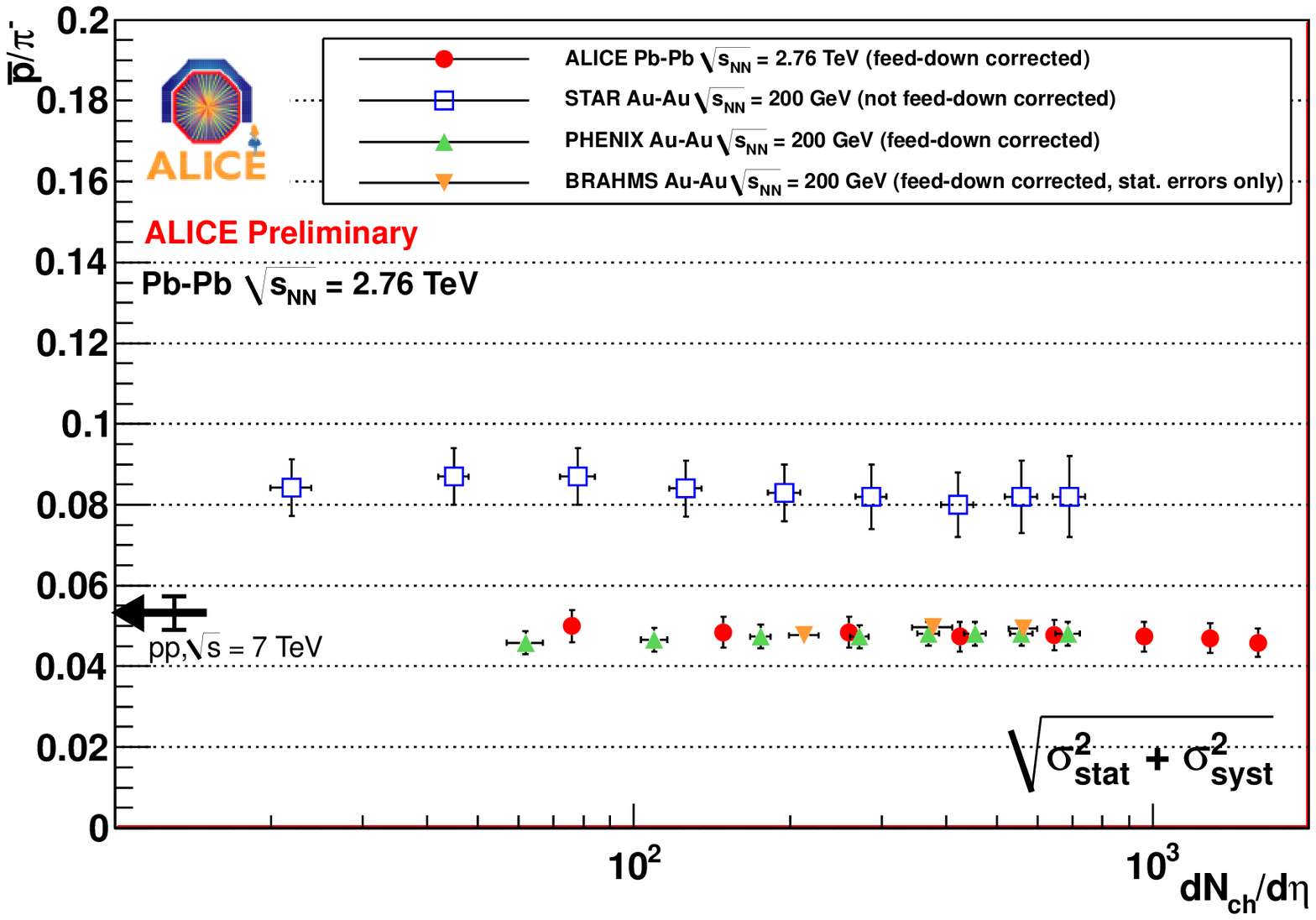}
 \caption{\kpim\ (left panel) and \ppim\ (right) ratios as a function of \dNdeta.}
  \label{fig:RatiosPbPb}
\end{figure}

The \pt-integrated \kpim\ and \ppim\ ratios as a function of \dNdeta\ are compared in Fig.~\ref{fig:RatiosPbPb} to RHIC ratios and to our pp measurement. The \kpim\ follows nicely the trend from lower energies. The \ppim\ ratio is similar to previous measurements by PHENIX and BRAHMS, which used a similar feed-down correction. The \ppim\ ratio measured by STAR is higher by a factor $\sim 1.5$, partially due to the fact that the \pbar\ from STAR are not corrected for feed-down. However, even when the feed-down correction at STAR is taken into account, some disagreement between the RHIC experiments persists, as it is apparent from the thermal analyses of the RHIC data (see e.g.~\cite{Anton-QM}). The observed \ppim\ ratio at the LHC is much lower than expected from thermal model analyses which, based on lower energy data, used the values $\mathrm{T}\simeq160-170~\mathrm{MeV}$ and $\mu_{b}\simeq 1~\mathrm{MeV}$ at the LHC, leading to \ppim$\simeq0.07-0.09$~\cite{Cleymans:2006xj,Andronic:2005yp}. The \kpim\ ratio, on the other hand, is in line with the expectations from those analyses.

\section{Conclusions}
\label{sec:conclusions}

We presented new measurements of identified particles made with the ALICE detector in pp and Pb-Pb collisions at LHC energies, which demonstrate the excellent PID capabilities of the experiment.
In pp collisions, the results at \s~=~900~GeV and 7~TeV show that most particle ratios are independent of energy in the TeV energy region.
In Pb-Pb collisions, the spectral shapes show stronger radial flow than at RHIC (about 10\% higher \betaT\ according to blast wave fits). The baryon/meson anomaly was investigated with the $\Lambda/\kzero$ ratio, and the enhancement is slightly higher and pushed towards higher \pt\ than at RHIC. The ratio p/$\pi$ is found to be about 0.05 both in pp and Pb-Pb collisions. This value is difficult to understand in a thermal model with T~=~$160-170$~MeV.

\section*{References}

\bibliographystyle{jphysg} 
\bibliography{mfloris_qm2011}

\end{document}